\documentstyle[12pt,preprint,aps,prb]{revtex}
\def\nm{\nonumber}
\def\Del{\Delta}
\def\beqa{\begin{eqnarray}}
\def\beq{\begin{equation}}  
\def\F{{\cal{F}}}
\def\L{{\cal{L}}}
\def\eeqa{\end{eqnarray}}
\def\eeq{\end{equation}}
\def\lab{\label}   
\def\pa{\partial}
\def\tx{\theta_{x}}
\def\ty{\theta_{y}}
\def\tz{\theta_{z}}
\def\tu{\theta_{u}}
\def\tv{\theta_{v}}
\def\tw{\theta_{w}}
\def\l{\Lambda}
\def\n{\nu_{1}}
\def\nn{\nu_{2}}
\def\nnn{\nu_{3}}
\def\ex{\epsilon_{x}}
\def\ey{\epsilon_{y}}
\def\ez{\epsilon_{z}}

\begin{document}

\begin{titlepage}
\thispagestyle{plain}
\pagenumbering{arabic}
\vspace*{-1.9cm}
\vspace{1.0cm}
\begin{center}
{\Large \bf One-Instanton Prepotentials from WDVV Equations in}
\end{center}
\vspace{-7.0mm}
\begin{center}
{\Large \bf $N=2$ Supersymmetric SU(4) Yang-Mills Theory}
\end{center}
\vspace{-7.0mm}
\lineskip .80em
\vskip 4em
\normalsize
\begin{center}
{\large Y\H uji Ohta}
\end{center}
\vskip 1.5em
\begin{center}
{\em Research Institute for Mathematical Sciences }
\end{center}
\vspace{-10.5mm}
\begin{center}
{\em Kyoto University}
\end{center}
\vspace{-10.5mm}
\begin{center}
{\em Sakyoku, Kyoto 606, Japan.}
\end{center}
\vspace{1.0cm}
\begin{abstract}
Prepotentials in $N=2$ supersymmetric Yang-Mills theories are known 
to obey non-linear partial differential equations called 
Witten-Dijkgraaf-Verlinde-Verlinde (WDVV) equations. In this paper, 
the prepotentials at one-instanton level in 
$N=2$ supersymmetric SU(4) Yang-Mills theory are studied 
from the standpoint of WDVV equations. Especially, it is shown that 
the one-instanton prepotentials are obtained from WDVV equations 
by assuming the perturbative prepotential and by using the scaling 
relation as a subsidiary condition but are determined without 
introducing Seiberg-Witten curve. In this way, various 
one-instanton prepotentials which satisfy both WDVV equations and 
scaling relation can be derived, but it turns out that among them there 
exist one-instanton prepotentials which coincide with the instanton 
calculus. 
\end{abstract}
\end{titlepage}


\begin{center}
\section{Introduction}
\end{center}

\renewcommand{\theequation}{1.\arabic{equation}}\setcounter{equation}{0}

A class of developments of quantum field theory in the ninety of this 
century may be represented by two keywords: $N=2$ supersymmetry and 
duality. For example, the mirror symmetry \cite{Yau,M1,Can,Can2} 
established in the beginning of 
the ninety was based on the (trivial) isomorphism between left and 
right $U(1)$ currents of $(2,2)$-superconformal field theory,\cite{Dix,GP2} 
and this isomorphism predicted the existence of a pair of 
Calabi-Yau manifolds whose axes of Hodge diamond were exchanged. 
Candelas {\em et al.} \cite{Can,Can2} skillfully used the consequence 
expected from this duality of Hodge structure and showed that 
the numbers of rational curves on Calabi-Yau quintic 3-fold could be 
determined from mirror symmetry. The coincidence of their result 
with mathematically rigorous results \cite{Yau,M2,Kat} 
gave a great surprising! 

On the other hand, also in the recent studies of 
low energy effective dynamics of $N=2$ 
supersymmetric Yang-Mills theory, $N=2$ supersymmetry and duality 
play a crucial role. Before the arrival of Seiberg and Witten's 
proposal by using electro-magnetic duality 
for the description of the low energy effective action of 
SU(2) gauge theory, \cite{SW1,SW2} 
though it has been known that the prepotential which is a 
generating function of the low energy effective action is not 
renormalized beyond one-loop in perturbative calculation 
due to $N=2$ supersymmetry, \cite{HST1,HST2,HSW} 
actually the prepotential was expected to receive instanton 
corrections. \cite{Sei} Unfortunately, such corrections were not so 
extensively discussed, but thanks to their proposal, it made possible to 
extract informations on instanton effects from 
a Riemann surface and periods of meromorphic differential on it. 
Namely, the low energy effective theory was turned out to be 
parameterized by a Riemann surface. The validity of their proposal 
was discussed by Klemm {\em et al.} \cite{KLT} with the aid of 
Picard-Fuchs equation and the instanton corrections to the prepotential 
was revealed. The instanton corrections obtained in this way 
showed extremely good agreement with the prediction of instanton 
calculus. \cite{FP,IS3,IS,Slat,DKM1,AHSW}

However, deeper and striking features of prepotentials of $N=2$ 
supersymmetric Yang-Mills theories may be nicely 
interpreted in terms of differential equations satisfied by prepotentials. 
For instance, it is well-known that the prepotentials satisfy an 
Euler equation called scaling relation, \cite{Mat,STY,EY,HW,KO} 
and in fact this simple equation simplified and accelerated the study of 
prepotentials. As for another characteristic equations, we can mention 
that there is a non-linear system of partial differential equations called 
Witten-Dijkgraaf-Verlinde-Verlinde 
(WDVV) equations \cite{MMM1,MMM2,MMM3,IY,IXY} 
(rigorously speaking, the WDVV equations in $N=2$ 
Yang-Mills theory are not equivalent to those arising in two-dimensional 
topological field theory \cite{Wit,DW,DVV,Dub}). Actually, these equations hold 
not only in four-dimensional gauge theories but also in higher dimensions 
even if hypermultiplets are included. \cite{MMM2,MMM3} 
Accordingly, it becomes possible to regard the prepotentials in various 
gauge theories as a member of solutions to WDVV equations. Then, what is 
the most general solution (function form of prepotentials) to the WDVV 
equations? Unfortunately, we can not precisely know the answer to 
this question, but Braden {\em et al.} \cite{BMMM} partially found 
the answer. They assumed the function form of prepotential which 
is expected from known examples and found a new prepotential which is 
considered as that in five-dimensional gauge theory, although their 
study was restricted to perturbative part. Of course, among the solutions 
found by them we can see the existence of the prepotential in 
four-dimensional Yang-Mills theory. This seems to indicate that the 
prepotentials can be constructed without introducing Riemann surface, 
provided the WDVV equations are used. Finding whether 
non-perturbative prepotentials are available from WDVV equations without 
using Riemann surface is the subject of this paper. 

The paper is organized as follows. In Sec. II, the construction of 
perturbative solution to WDVV equations for SU(4) gauge theory in four 
dimensions discussed by Braden {\em et al.} \cite{BMMM} is summarized. 
We can see that the perturbative prepotential is in fact obtained from 
WDVV equations. In Sec. III, we add the non-perturbative part for this 
perturbative prepotential and try to solve the WDVV equations. 
Though the non-perturbative part satisfies a non-linear differential 
equation, restricting it at one-instanton level, we can reduce it to a 
linear differential equation satisfied by one-instanton prepotential. 
For this reason, the one-instanton prepotential is investigated in this 
paper. To solve this equation, the scaling relation is used as a subsidiary 
condition, but it turns out that there are miscellaneous solutions which 
do not contradict to both WDVV equations and scaling relation. In 
Sec. IV. we compare our result with the prediction of one-instanton 
calculus. It is shown that among our one-instanton 
prepotentials obtained from WDVV equations there are 
one-instanton prepotentials which agree the prediction of 
instanton calculus. In this way, we conclude that it is possible to 
obtain non-perturbative prepotential from WDVV equations without 
relying on Riemann surface. Sec. V is a brief summary. 

\begin{center}
\section{Perturbative prepotential from WDVV equations }
\end{center}

\renewcommand{\theequation}{2.\arabic{equation}}\setcounter{equation}{0}

In this section, we briefly outline the construction to get 
perturbative prepotential from WDVV equations in the SU(4) gauge theory 
presented by Braden {\em et al}. \cite{BMMM} Note that the SU(4) model 
is the simplest and non-trivial example for a study of WDVV equations.

In this case, the WDVV equations for the prepotential $\F$ take the form  
	\beq
	(\F_i )(\F_k )^{-1}(\F_j )=(\F_j )(\F_k )^{-1}(\F_i )
	,\ i,j,k =1,\cdots,3
	,\lab{wdvv}
	\eeq
where 
	\beq
	(\F_i )\equiv (\F_i )_{jk}=\frac{\pa^3 \F}{\pa a_i \pa a_j 
	\pa a_k}
	\eeq
are the matrix notations and in this paper the brackets are always added 
as $(\F_i )$ when ``$\F_i$'' mean matrices. The coordinates $a_i$ are 
the periods of the SU(4) gauge theory (see Appendix). 

Braden {\em et al.} \cite{BMMM} considered the perturbative prepotential 
in the form 
	\beq
	\F_{\mbox{\scriptsize per}}(a_1 ,a_2 ,a_3 ) =
	\sum_{i<j=1}^{4}f(a_{ij}),\ a_{ij}=a_i -a_j ,
	\ \sum_{i=1}^4 a_i =0
	.\lab{22}
	\eeq
Of course $\F_{\mbox{\scriptsize per}}$ may depend on the mass scale 
$\l \equiv \l_{\mbox{\scriptsize SU(4)}}^8$ of the 
theory, but we can ignore its dependence for the moment 
because $\l$-differentiation is 
not included in the WDVV equations (\ref{wdvv}). 

Under the assumption (\ref{22}), we can find 
that when $\F_{\mbox{\scriptsize per}}$ 
satisfies (\ref{wdvv}) there is a functional relation 
	\beq
	g(a_{12})g(a_{34})-g(a_{13})g(a_{24})+g(a_{14})g(a_{23})=0
	,\lab{23}
	\eeq
where
	\beq
	g(a)\equiv \left(\frac{\pa^3 f}{\pa a^3}\right)^{-1}
	.\lab{ga}
	\eeq
With the aid of several conditions, we can conclude 
that $g$ is an odd function with 
	\beq
	g(0)=g^{\,''}(0)=0,
	\lab{226}
	\eeq
where the prime means the differentiation over the 
argument. \cite{BMMM} 

There are several functions enjoying the properties (\ref{23}) and 
(\ref{226}), but a function which is necessary for us among them is 
the function of the form $g(a)=a$. Namely,  
	\beq
	f(a)=\frac{a^2}{2}\ln a+O(a^2)
	.\lab{26}
	\eeq
Note that $O(a^2)$-term can not be fixed from the WDVV equations because 
they are third-order differential equations. Namely, 
$\l$-dependence of one-loop contribution is not fixed. It is easy to 
see that substituting (\ref{26}) back to (\ref{22}) in fact yields the 
perturbative part of the 
SU(4) prepotential (in a suitable normalization). 

Marshakov {\em et al.} \cite{MMM3} give a general proof that the 
perturbative prepotentials in various gauge theories satisfy WDVV 
equations, but this is also confirmed by Ito and Yang \cite{IY} in their 
study of these equations. 

\begin{center}
\section{One-instanton prepotentials}
\end{center}

\renewcommand{\theequation}{3.\arabic{equation}}\setcounter{equation}{0}

\begin{center}
\subsection{Differential equation for one-instanton prepotential}
\end{center}


Next, let us consider whether non-perturbative prepotential $\F$ 
is available from the WDVV equations by assuming the form  
	\beq
	\F (a_1 ,a_2 ,a_3 ,\l )=\F_{\mbox{\scriptsize per}}(a_1 ,a_2 ,a_3 )
	 +\F_{\mbox{\scriptsize ins}}(a_1 ,a_2 ,a_3 ,\l )
	,\lab{2}
	\eeq
where 
	\beq
	\F_{\mbox{\scriptsize ins}}(a_1 ,a_2 ,a_3 ,\l )=
	\sum_{k=1}^{\infty}\F_k (a_1 ,a_2 ,a_3 ) \l^k 
	.\eeq
In order to derive differential equations 
for $\F_{\mbox{\scriptsize ins}}$, we assume that 
$\F_{\mbox{\scriptsize per}}$ is already given by (\ref{22}) 
with (\ref{26}). 

Then substituting $\F$ into (\ref{wdvv}) we can obtain a single 
non-linear differential equation for $\F_{\mbox{\scriptsize ins}}$, 
but if we restrict only the case $k=1$ (one-instanton level), 
the equation reduces to 
	\beqa
	& &\frac{\pa_{1}^3 \F_1}{a_{12}a_{13}a_{14}}-\frac{\pa_{2}^3 \F_1}
	{a_{12}a_{23}a_{24}}+\frac{\pa_{3}^3 \F_1}{a_{13}a_{23}a_{34}}
	+\frac{6\pa_1\pa_2\pa_3\F_1}{a_{14}a_{24}a_{34}}
	-\frac{A_{012}\pa_2\pa_{3}^2\F_1 +A_{021}\pa_{2}^2\pa_3\F_1}
	{a_{12}a_{13}a_{23}a_{24}a_{34}}\nm\\
	& &-\frac{A_{102}\pa_1\pa_{3}^2 \F_1 
	-A_{201}\pa_{1}^2\pa_3 \F_1}{a_{12}a_{13}a_{14}a_{23}a_{34}}
	+\frac{A_{120}\pa_1\pa_{2}^2\F_1 +A_{210}\pa_{1}^2\pa_2\F_1}
	{a_{12}a_{13}a_{14}a_{23}a_{24}}=0
	\lab{k}
	,\eeqa
where $\pa_i \equiv \pa /\pa a_i$ and 
	\beqa
	& &A_{012}=a_1 a_2 -3a_{2}^2 -2a_1 a_3 +4a_2 a_3+a_1 a_4 +a_2 a_4
	-2 a_3 a_4 ,\nm\\
	& &A_{021}=2a_1 a_2 -a_1 a_3 -4a_2 a_3 +3a_{3}^2 -a_1 a_4 +2
	a_2 a_4 -a_3 a_4 ,\nm\\
	& &A_{102}=3a_{1}^2 -a_1 a_2 -4a_1 a_3 +2a_2 a_3 -a_1 a_4 -a_2 a_4 
	+2a_3 a_4 ,\nm\\
	& &A_{201}=2a_1 a_2 -4a_1 a_3 -a_2 a_3 +3a_{3}^2 +2a_1 a_4 -a_2 a_4 
	-a_3 a_4 ,\nm\\
	& &A_{120}=3a_{1}^2 -4a_1 a_2 -a_1 a_3 +2a_2 a_3 -a_1 a_4 +2a_2 a_4 
	-a_3 a_4 ,\nm\\
	& &A_{210}=4a_1 a_2 -3a_{2}^2 -2a_1 a_3 +a_2 a_3 -2a_1 a_4 +
	a_2 a_4 +a_3 a_4
	.\eeqa
	
\begin{center}
\subsection{The solutions}
\end{center}

In order to solve (\ref{k}), let us introduce the new variables
	\beq
	x=a_{12} ,\ y=a_{13},\ z=a_{14}
	\lab{111}
	.\eeq
In addition, using Euler derivatives $\tx =x\pa /\pa x$ etc, we 
can rewrite (\ref{k}) as 
	\beq
	L(\tx ,\ty ,\tz )\F_1 =0
	,\eeq
where 
	\beqa
	L(\tx ,\ty ,\tz )&=&yz(y-z)\tx (\tx -1)(\tx -2)+z(4xy-3y^2 -2xz+yz)
	(\tx -1)\tx \ty \nm\\
	& &+z(3x^2 -4xy-xz +2yz )(\ty -1)\tx \ty -xz(x-z)\ty (\ty-1)(\ty -2)
	\nm\\
	& &-y(3x^2 -xy-4xz +2yz)(\tz -1)\tx \ty 
	-y(4xz+yz -3z^2 -2xy )(\tx -1)\tx \tz \nm\\
	& &+6(x-y)(x-z)(y-z)\tx \ty \tz +
	x(xz+4yz -3z^2 -2xy)(\ty -1)\ty \tz \nm\\
	& &+x(3y^2 +2xz -4yz -xy)(\tz -1)\ty \tz +
	xy(x-y)\tz (\tz -1)(\tz -2)
	.\eeqa
Here, suppose that $\F_1$ is given by 
	\beq
	\F_1 =x^{\nu_1}y^{\nu_2}z^{\nu_3}F(x,y,z)
	\lab{nu}
	,\eeq
where 
	\beq
	F(x,y,z)=
	\sum_{i,j,k=0}^{\infty}B_{\ex i,\ey j,\ez k}x^{\ex i} y^{\ey j}
	z^{\ez k}
	\lab{b}
	\eeq
and the expansion coefficients are assumed to be independent 
of $x,y$ and $z$. In (\ref{b}), $\ex$ etc are signature symbols 
(the choice of signatures depends 
on where is the convergence region), 
e.g., $\ex =\pm$. Then from (\ref{k}) we get the differential 
equation for $F$
	\beq
	L (\tx +\n ,\ty +\nn ,\tz +\nnn )F =0
	.\lab{cc}
	\eeq
and the indicial equations for $\nu_i$ 
	\beqa
	& &\n (\n -2\nn -1)(\n +\nn -3\nnn -2)=0,\ 
	\nn (2\n -\nn +1)(\n +\nn -3\nnn -2)=0,\nm\\
	& &\nn (\nn -2\nnn -1)(3\n -\nn -\nnn +2)=0,\ 
	(\nn -\nnn )(\n^2 -\n -\n\nn +\nn \nnn )=0,\nm\\
	& &\n (\n^2 -3\n +5\nn -3\n\nn +\nnn -\n\nnn +4\nn\nnn +2)=0
	,\nm\\
	& &\nnn^2 (\nnn -3\nn-3)+5\n\nn\nnn +3\nn\nnn -2\n^2 \nnn+
	2\n\nnn+2\nnn +\n\nn =0,\nm\\
	& &\nnn^2 (\nnn -\nn-3)-2\nn^2 \nnn +3\n\nn\nnn +3\nn\nnn +3\n\nn +
	2\nnn =0
	.\eeqa
The sets of possible $\nu_i$ are thus given by 
	\beqa
	\mbox{\boldmath$\nu$}\equiv (\n,\nn,\nnn)&=&(-2,-2,-2),\ 
	(-1,-1,-1),\ (0,-1,-1)^4 ,\ (0,0,0)^2 ,\ (0,0,1),\ (0,0,2)^2 ,\nm\\
	& & (0,1,0),\ (0,1,1),\ (0,2,0)^2 ,\ (1,0,0)^2 ,\ 
	(1,0,1),\ (1,0,2),\ (2,0,0)^2 
	,\lab{list}
	\eeqa
where the superscript means degeneracy, e.g., $(0,0,0)^2$ 
is composed of two $(0,0,0)$, but we do not discuss the consequence of 
degeneracy in this paper. 
For the indices (\ref{list}), it is straightforward to obtain $F$. Though 
we have expressed $F$ in (\ref{b}) as an infinite series, 
actually we can restrict possible terms in (\ref{b}) by considering 
the degree counting of one-instanton prepotential. 

\begin{center}
\subsection{Scaling relation for one-instanton prepotential} 
\end{center}

If the WDVV equations can in fact yield a physically acceptable 
prepotential, the prepotential obtained from those equations must also 
satisfy the fundamental homogeneity condition called scaling 
relation. \cite{Mat,STY,EY,HW,KO} Therefore, we may use it as a 
subsidiary condition for the problem how to solve WDVV equations in 
gauge theory. 

To see this, firstly, let us recall the scaling 
relation \cite{Mat,STY,EY,HW,KO} 
	\beq
	\sum_{i=1}^{3}a_i \frac{\pa \F}{\pa a_i} +
	\l_{\mbox{\scriptsize SU(4)}} \frac{\pa \F}
	{\pa \l_{\mbox{\scriptsize SU(4)}}}=2\F
	.\lab{sca}
	\eeq
We need a scaling relation for $\F_1$ not for $\F$ itself, but 
in order to extract it from (\ref{sca}), the 
$\l$-dependence of perturbative prepotential which can not be fixed 
from WDVV equations must be included appropriately. For this, our choice 
here is 
	\beq
	\F_{\mbox{\scriptsize per}}=\sum_{i<j=1}^{4}\frac{a_{ij}^2}{2}
	\ln \frac{a_{ij}}{\l_{\mbox{\scriptsize SU(4)}}}
	.\eeq
Then from (\ref{sca}), $\F_1$ is found to satisfy 
	\beq
	\sum_{i=1}^3 a_i \frac{\pa \F_1}{\pa a_i}
	+6\F_1 =0
	\lab{42}
	,\eeq
which indicates that $\F_1$ is a homogeneous function of degree $-6$. 

In the variables (\ref{111}), (\ref{42}) becomes 
	\beq
	x\pa_x \F_1 +y\pa_y \F_1 +z\pa_z \F_1 +6\F_1 =0
	.\lab{318}
	\eeq
Accordingly, from (\ref{nu}), (\ref{b}) and (\ref{318}) 
it must be always true that 
	\beq
	\n +\nn +\nnn +\ex i+\ey j+\ez k =-6
	.\lab{319}
	\eeq

\begin{center}
\subsection{Examples of one-instanton prepotentials} 
\end{center}

We have now enough informations to construct explicit one-instanton 
prepotentials which do not contradict to WDVV equations and 
scaling relation. 

To begin with, let us consider the 
case $\mbox{\boldmath$\epsilon$}\equiv (\ex ,\ey ,\ez)=(+,+,+)$. 
In this case, 
we can easily find that there exists only one solution which satisfies 
(\ref{319}). It is the solution with $\mbox{\boldmath$\nu$}=(-2,-2,-2)$, 
and thus  
	\beq
	\F_1 =\frac{B_{0,0,0}}{x^2 y^2 z^2 }
	.\lab{f1}
	\eeq

However, when $\mbox{\boldmath$\epsilon$}=(-,-,-)$ or one entry of 
$\mbox{\boldmath$\epsilon$}$ differs to the others, e.g., 
$\mbox{\boldmath$\epsilon$} =(-,-,+)$, the situation changes, 
in particular, drastically in the latter case. In the former case, 
it is easy to see that $F$ consists of 
finite number of terms for all indices in (\ref{list}), 
but in the latter case $F$ is generally represented by 
infinite number of terms as long as (\ref{319}) is satisfied. 
We do not know whether it is possible to 
find any physical meaning for this type of one-instanton prepotential, but 
it may be interesting to recall that a 
similar one was observed in the 
one-instanton prepotential in the five-dimensional gauge theory.\cite{KO} 

Since the latter case mentioned above is slightly intractable, let us 
consider an example of the former case instead. In the case 
of $\mbox{\boldmath$\nu$}=(-1,-1,-1)$ 
with $\mbox{\boldmath$\epsilon$}=(-,-,-)$, for instance, we have  
	\beqa
	\F_1 &=&\frac{1}{xyz}
	\left[B_{0,-1,-2}\left(\frac{1}{yz^2}+\frac{1}{y^2 z}\right)
	+B_{-1,0,-2}\left(\frac{1}{xz^2}+\frac{1}{x^2 z}\right)\right.\nm\\
	& &\left. +B_{-1,-2,0}\left(
	\frac{1}{x^2 y}+\frac{1}{xy^2}\right) +B_{-1,-1,-1}\frac{1}{xyz}
	\right]
	.\lab{3333}
	\eeqa
Note that (\ref{3333}) includes the 
one-instanton prepotential of the form (\ref{f1}). 

In a similar manner, we can construct 
one-instanton prepotentials for all other possible values 
of $\nu_i$, which do not contradict to both WDVV equations and scaling 
relation, but it would not be necessary to explicitly show them here. 
However, we should point out that since (\ref{k}) is 
a partial differential system we can expect that there exist more 
and more various solutions. In fact, this observation is right, and 
we can show that also in the variables  
	\beq
	(x,y,z)=(a_{12},a_{23},a_{24}),\ (a_{13}, a_{23},a_{34})
	\lab{lat}
	\eeq
we can construct miscellaneous one-instanton prepotentials. 
Among them one-instanton 
prepotentials of the form (\ref{f1}) are included. 

{\bf Remark:} {\em The function form of the one-instanton prepotentials 
can be determined by solving the WDVV equations, but its numerical factors, 
i.e., instanton expansion coefficients, are not obtained because 
they correspond to integration constants. In order to determine them, it 
is necessary to rewrite the scaling relation as a relation between prepotential 
and moduli. Then substituting 
the one-instanton prepotential obtained from WDVV equations into this 
scaling relation, we will be able to get the expansion coefficients. 
Of course, in this case the moduli must be represented as a 
function of periods and its expansion coefficients must be determined. 
However, since knowing moduli is equivalent to 
introduce a Seiberg-Witten curve, this method based on scaling relation 
represented by using moduli is not preferable in the 
formalism of WDVV equations because prepotentials available from 
WDVV equations should be determined without introduction of 
Seiberg-Witten curves. Accordingly, when the determination of instanton 
expansion coefficients is required, they should be determined from 
the result of instanton calculus.}

\begin{center}
\section{One-instanton prepotential from instanton calculus} 
\end{center}

\renewcommand{\theequation}{4.\arabic{equation}}\setcounter{equation}{0}

We have derived one-instanton prepotentials by solving the WDVV equations 
in the previous section. Though these one-instanton prepotentials satisfies 
the WDVV equations and the scaling relation, 
unfortunately in view of WDVV equations we can not determine which ones 
are physically acceptable. For this reason, in order to extract 
physically meaningful one-instanton prepotentials among them, 
we must compare our result with the one-instanton prediction of instanton 
calculus. 

In the case of SU(4) gauge theory, one-instanton contribution for 
prepotential is given by \cite{IS3,IS} 
	\beq
	\F_1 =\frac{\Del_{4}^{\,'}}{\Del_4}
	,\lab{37}
	\eeq
where we have omitted the numerical normalization factor and  
	\beq
	\Del_{4}^{\,'}=\sum_{i=1}^{4}
	\prod_{\stackrel{k<l=1}{k,l\neq i}}^{4}(a_k -a_l )^2 ,\ 
	\Del_4 =\prod_{k<l=1}^{4}(a_k -a_l )^2 
	.\eeq
The closed form of one-instanton prepotential for SU($N_c$) gauge theory 
is also obtained by solving 
Picard-Fuchs equations \cite{IY2} and direct calculation of period 
integrals. \cite{DKP,MS}

Note that (\ref{37}) is a sum of (\ref{f1}) and those for (\ref{lat}) 
	\beq
	\F_1 =\frac{1}{(a_{12}a_{13}a_{14})^2}+\frac{1}{
	(a_{12}a_{23}a_{24})^2}+\frac{1}{(a_{13}a_{23}a_{34})^2}
	\lab{310}
	\eeq
up to constant factors. Accordingly, we can conclude that the WDVV 
equations can yield physical prepotential in spite of without introducing 
Riemann surface. 

\begin{center}
\section{Summary}
\end{center}

\renewcommand{\theequation}{4.\arabic{equation}}\setcounter{equation}{0}

In this paper, we have discussed the non-perturbative prepotential of 
$N=2$ supersymmetric SU(4) Yang-Mills theory in the standpoint of WDVV 
equations. Especially, we have found a differential equation for 
one-instanton prepotential and constructed its solutions. 
The method to get prepotentials based on WDVV equations is fascinating 
in the point that the prepotentials can be obtained without introducing 
Seiberg-Witten curves, but it has been shown that 
unfortunately too many prepotentials exist in 
contrast with the approach based on Seiberg-Witten curves 
which uniquely determines a prepotential. Nevertheless, 
we have succeeded to show that one-instanton prepotentials which coincide 
with the one-instanton calculus can be obtained from WDVV equations.

As for an another aspect of WDVV equations, we should mention a connection 
to topological field theory in two dimensions. 
From the appearance of WDVV equations, it may be natural to think that 
the low energy effective theory is actually a kind of topological field 
theory, but we must notice that a priori there is no reason that the 
effective theory must be a topological field theory. 
Therefore, topological or not topological: that is the question. 

An approach to argue this implication more explicitly is to 
regard the Seiberg-Witten curves often identified with spectral curves 
of integrable system as if they were superpotentials of 
topological $\mbox{\boldmath$CP$}^1$ model. \cite{IXY} Although this observation 
strongly relying on the existence of Riemann surface (Seiberg-Witten curve), 
it enables us to 
find a connection to topological field theory, specifically, 
topological string theory at genus zero level. \cite{IXY} Then, even if we 
do not assume a Riemann surface, can we find a topological nature of 
the effective theory? Probably WDVV equations give the answer, 
but the study is a subject in the future. 

\begin{center}
\section*{Acknowledgment}
\end{center}

The author acknowledges Prof. H. Kanno for discussions about 
WDVV equations. 

\begin{center}
\section*{Appendix: The SU(4) Seiberg-Witten solution}
\end{center}

\renewcommand{\theequation}{A\arabic{equation}}\setcounter{equation}{0}

In this appendix, we briefly summarize the SU(4) Seiberg-Witten solution. 
The SU(4) Seiberg-Witten curve is given by the hyperelliptic curve of 
genus three \cite{KLYT,AF,HO,AAM}
	\beq
	y^2 =(x^4 -ux^2 -vx-w)^2 -\l_{\mbox{\scriptsize SU(4)}}^8
	,\eeq
where $(x,y)\in \mbox{\boldmath$C$}^2$ is the local coordinate, 
$u,v$ and $w$ are moduli of the theory. Then the Seiberg-Witten 
differential and its periods are given by 
	\beq
	\lambda_{\mbox{\scriptsize SW}}=\frac{x\pa_x W}{y}dx
	\eeq
and 	
	\beq
	a_i =\oint_{\alpha_i}\lambda_{\mbox{\scriptsize SW}},\ 
	a_{D_i}=\oint_{\beta_i}\lambda_{\mbox{\scriptsize SW}}
	,\ i=1,2,3 
	,\eeq
respectively, where $\alpha_i$ and $\beta_i$ are the canonical bases 
of the 1-cycles on the curve and the numerical normalization factor 
of the Seiberg-Witten differential is ignored. It is convenient to 
use the period vector 
	\beq
	\Pi =\left(\begin{array}{c}
	a_{D_i}\\
	a_i \end{array}\right)
	.\eeq

In general, these periods satisfy Fuchsian differential equations and 
in the case at hand they are given by \cite{KLT,IS,IMNS,Ali} 
	\beqa
	\L_1 \Pi &\equiv& \left[ \pa_{v}^2 -\pa_u \pa_w\right]\Pi=0,\nm\\
	\L_2 \Pi &\equiv& \left[ 4\pa_{u}^2 -2u\pa_u \pa_w -
	v\pa_v\pa_w -\pa_w \right]\Pi=0,\nm\\
	\L_3 \Pi &\equiv& \left[ v\pa_{w}^2 +2u\pa_v\pa_w -4\pa_u\pa_v 
	\right]\Pi=0,\nm\\
	& &\left[ 4(u^2 +24w)\pa_{u}^2 +9v^2\pa_{v}^2 -16 
	(\l_{\mbox{\scriptsize SU(4)}}^8 -w^2 )\pa_{w}^2 
	+12uv\pa_u\pa_v \right.\nm\\
	& &\left.-32uw\pa_u\pa_w +3v\pa_v -16w\pa_w +1 \right]\Pi=0
	,\lab{imns1}
	\eeqa
which can be summarized as  
	\beqa
	& &\left[\tv (\tv -1) -\frac{v^2}{uw}\tu\tw \right]\Pi =0,\nm\\
	& &\left[(2\tu +\tv +1)\tw -\frac{4w}{u^2}\tu (\tu -1)
	\right]\Pi =0,\nm\\
	& &\left[(2\tu +3\tv +4\tw -1)^2 -\frac{16\l_{\mbox{\scriptsize 
	SU(4)}}^8}{w^2}\tw (\tw -1)\right]\Pi =0
	,\lab{ne}
	\eeqa
where $\tw =w\pa_w$ etc are Euler derivatives, 
provided the first, second and last equations in (\ref{imns1}) are 
chosen as independent equations. Note that the third one in (\ref{imns1}) 
is not an independent equation 
because $(v\pa_w \L_1 +\pa_v \L_2 +\pa_u \L_3 )\Pi =0$. 

Introducing new variables $x,y$ and $z$ by
	\beq
	x=\frac{\l_{\mbox{\scriptsize SU(4)}}^8}{4w^2},\ 
	y=\frac{v^2}{4uw},\ z=\frac{w}{u^2}
	,\eeq
we find that (\ref{ne}) is converted into  
	\beqa
	& &\left[(8\tx +1)^2 -64x(2\tx+\ty-\tz)(2\tx+\ty-\tz +1)
	\right]\Pi =0,\nm\\
	& &\left[\ty (2\ty -1) -2y(\ty+2\tz)
	(2\tx+\ty -\tz)\right]\Pi =0,\nm\\
	& &\left[(2\tx+\ty-\tz)(4\tz -1)-4z (\ty +2\tz)(\ty+2\tz +1)
	\right]\Pi =0
	.\lab{SU4PF}
	\eeqa
This system (\ref{SU4PF}) further simplifies to 
	\beqa
	& &\left[\tx^2 -x\left(2\tx+\ty-\tz-\frac{1}{2}\right)
	\left(2\tx+\ty-\tz +\frac{1}{2}\right)\right]\widetilde{\Pi}=0,
	\nm\\
	& &\left[\ty \left(\ty -\frac{1}{2}\right) -
	y\left(\ty+2\tz+\frac{1}{2}\right)
	\left(2\tx+\ty -\tz-\frac{1}{2}\right)\right]\widetilde{\Pi}=0,
	\nm\\
	& &\left[\left(2\tx+\ty-\tz-\frac{1}{2}\right)
	\tz -z \left(\ty +2\tz +\frac{1}{2}\right)
	\left(\ty+2\tz +\frac{3}{2}\right)
	\right]\widetilde{\Pi}=0
	\eeqa
by $\Pi =x^{-1/8}z^{1/4}\widetilde{\Pi}$. An analytic solution 
around $(x,y,z)=(0,0,0)$ is given by 
	\beq
	\widetilde{\Pi}=\sum_{m,n,p =0}^{\infty}
	\frac{(1/2)_{n+2p}(-1/2)_{2m+n-p}}{(1)_m (1/2)_n}
	\frac{x^m}{m!}\frac{y^n}{n!}\frac{z^p}{p!}
	,\eeq
which is known as the type $54b$ 
Srivastava and Karlsson's (Gaussian) hypergeometric function in three 
variables, \cite{SK} and we denote it by 
	\beq
	G_{54b}[\alpha,\beta;\gamma,\delta;x,y,z]=
	\sum_{m,n,p =0}^{\infty}
	\frac{(\alpha)_{n+2p}(\beta)_{2m+n-p}}{(\gamma)_m (\delta)_n}
	\frac{x^m}{m!}\frac{y^n}{n!}\frac{z^p}{p!}
	,\lab{223}
	\eeq 
which recovers Horn's ${\cal{H}}_4$ for $p=0$. Note that $G_{54b}$ 
satisfies  
	\beqa
	& &\left[(\tx +\gamma -1)\tx -x\left(2\tx+\ty-\tz+\beta\right)
	\left(2\tx+\ty-\tz +\beta +1\right)\right]G_{54b}=0,\nm\\
	& &\left[(\ty +\delta -1)\ty -y\left(\ty+2\tz+\alpha \right)
	\left(2\tx+\ty -\tz +\beta\right)\right]G_{54b}=0,\nm\\
	& &\left[\left(2\tx+\ty-\tz+\beta\right)
	\tz -z \left(\ty +2\tz +\alpha \right)
	\left(\ty+2\tz +\alpha +1\right)
	\right]G_{54b}=0
	.\eeqa

\begin{center}

\end{center}

\end{document}